\documentclass[prl,aps,preprint,showpacs,preprintnumbers,amsmath,amssymb]{revtex4}

\usepackage{graphicx}
\usepackage{dcolumn}
\usepackage{bm}
\usepackage{natbib}
\usepackage{amsmath}
\usepackage{here}

\begin{document}

\preprint{preprint - not for distribution}

\title{Direct Measurement of the Spin-Orbit Interaction in a Two-Electron InAs Nanowire Quantum Dot}
\author{C. Fasth,$^{1}$ A. Fuhrer,$^{1}$ L. Samuelson,$^1$ Vitaly N. Golovach,$^{2}$ Daniel Loss,$^2$}

\affiliation{
$^1$Solid State Physics/Nanometer Consortium, Lund University, P.O. Box 118 Lund,  Sweden\\
$^2$Department of Physics and Astronomy, University of Basel, Klingenbergstrasse 82, CH-4056 Basel, Switzerland}

\date{\today}

\begin{abstract}
We demonstrate control of the electron number down to the last
electron in tunable few-electron quantum dots defined in catalytically
grown InAs nanowires. Using low temperature transport spectroscopy in
the Coulomb blockade regime we propose a simple method to directly
determine the magnitude of the spin-orbit interaction in a
two-electron artificial atom with strong spin-orbit coupling. Due to a
large effective g-factor $|g^*|=8\pm1$ the transition from singlet $S$
to triplet $T^+$ groundstate with increasing magnetic field is
dominated by the Zeeman energy rather than by orbital effects. We find
that the spin-orbit coupling mixes the $T^+$ and $S$ states and thus
induces an avoided crossing with magnitude
$\Delta_{SO}=0.25\pm0.05\;\mathrm{meV}$. This allows us to calculate
the spin-orbit length $\lambda_{SO}\approx127$ nm in such systems using a simple model.
\end{abstract} 

\pacs{Valid PACS appear here}
%\pacs{73.23.Hk, 73.63.Kv, 71.70.Ej}

\maketitle

Semiconducting nanowires are presently subject to an intense research
effort due to their potential as nano-scale building blocks for
components such as pn-junctions,\cite{Haraguchi92,Cui01} field effect
transistors,\cite{Greytak04,Goldberger05} logical
elements\cite{Huang01} and single electron
circuits.\cite{03thelander,DeFranceschi03,Lu05,Zhong05,Bjork04} For
spintronic or quantum electronic applications, e.g. qubits and quantum
gates,\cite{Loss98,Burkard99} accurate control over the
electron number down to the last one aswell as precise control over the spin state\cite{06koppens} and the coupling
between spins in neighboring quantum dots\cite{05petta} is
required. Furthermore, the spin-orbit interaction is of crucial
importance since coupling of the spin to orbital degrees of freedom and thus charge determines spin relaxation
and decoherence rates, which in turn determines decay rates for spin
qubits. 
It is thus of fundamental interest to experimentally determine the
spin-orbit length $\lambda_{SO}$ as accurately as possible.
The spin-orbit interaction is on the other hand also a pathway to directly manipulate the spin state of a quantum dot electrically rather than with conventional electron spin resonance techniques. In this context it was recently suggested to employ the spin-orbit coupling for all electrical spin-manipulation\cite{06golovach, 06flindt, 05debald} an approach which has the potential to be considerably faster than other schemes. Here, we discuss few-electron quantum dot formation in homogeneous InAs nanowires where the confined region is achieved by electrical depletion through local gate electrodes. This technique allows for highly tunable quantum dot systems with strong spin-orbit coupling where size, coupling and electron number starting from zero are easily controlled by external voltages and the spin-orbit length $\lambda_{SO}$ can be determined from a simple measurement of the anticrossing between singlet and triplet states.

%figure1
\begin{figure}[t!]
\includegraphics[width=6.3cm]{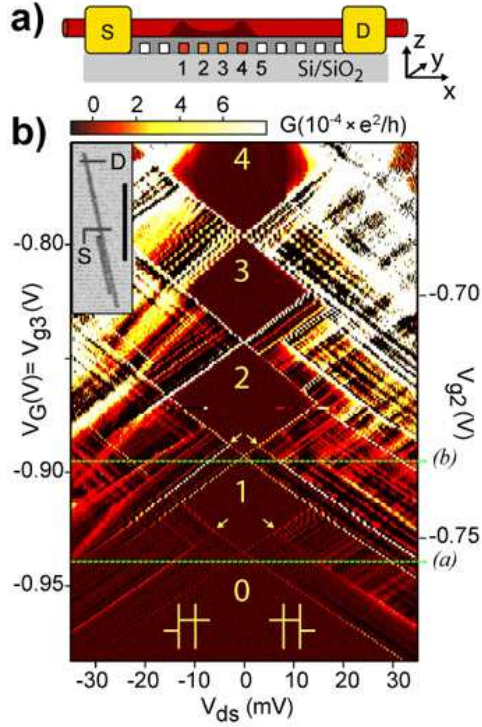}
\caption{\label{diamond}(a) Sample schematic showing the Au grid covered in SiN (dark grey) and the nanowire (red) on top. The source and drain contacts and the five individually contacted gates are indicated. (b) Differential conductance measured as a function of $V_{ds}$ and plunger gate voltages. The other gates are set to $V_{g1}=-2.56$V, $V_{g4}=-1.65$V and $V_{g5}=2.5$V, respectively. Electron numbers are indicated in yellow and the inset shows an SEM image of the wire on the electrode grid, with a 1 $\mu$m scalebar.}
\end{figure}

InAs nanowires were grown using chemical beam epitaxy with 40 nm Au catalyst particles, yielding untapered nanowires with a diameter of $\sim50\;$nm. For details on growth, see Ref. \onlinecite{Ohlsson02}. Electronic transport experiments show these wires to be n-type and have a mean free path $\ell\approx100\mathrm{nm}$.\cite{05hansen}
For the fabrication of gate induced quantum dots the wires are deposited onto a grid of gold electrodes, which is covered by a thin SiN dielectric layer. Using this technique we have previously demonstrated formation of gate-defined single and double many-electron quantum dots.\cite{Fasth05} Here, due to refined sample fabrication involving thinner nanowires, more densely spaced electrodes (periodicity 60 nm) and a thinner SiN layer (18 nm), the control over the potential landscape in the nanowire has been greatly improved, allowing tunability down to the last electron.

The wire and five gate electrodes are contacted individually with Ni/Au contacts [see Fig. \ref{diamond}(a) and inset to Fig. \ref{diamond}(b)]. The bias is applied symmetrically across the wire and measurements were performed in a dilution refrigerator at an electron temperature $T_e = 80$ mK. We apply negative voltages to gates 1 and 4 to locally deplete the electron density in the wire and form potential barriers, inducing a quantum dot above gates 2 and 3.  Figure \ref{diamond}(b) shows Coulomb blockade diamonds measured with gates 2 and 3 used as plunger gates. 
In the following we refer to the plunger gate voltage as
$V_{G}=V_{g3}$ and emphasize here that $V_{g2}$ is always tuned
simultaneously as in Fig.\;\ref{diamond}(b). Inside each diamond the
electron number $N$ on the dot is fixed, while at each diamond apex
close to zero bias the dot ground state energies for $N$ and $N+1$
electrons are degenerate and transport can occur. In
Fig.\;\ref{diamond}(b) no further degeneracy points are observed for
$V_{G}<-0.94$V and the diamond borderlines continue without kinks,
indicating that the quantum dot is empty\cite{suppl}. 
The diamond border-line conductance peaks correspond to the quantum
dot chemical potential being aligned to $\mu_{D}$ or $\mu_{S}$, as
indicated in Fig.\;\ref{diamond}(b). Additional peaks in the
differential conductance represent transitions involving excited states.

In addition to conductance peaks originating in dot states
Fig. \ref{diamond}(b) also shows an abundance of weak and densely
spaced peaks, which are related to the low-dimensional nature of the
source and drain leads\cite{suppl}. The unused local gates beneath the wire are
set to +2.2\;V in order to tune the leads towards metallic behavior
and avoid accidental dot formation. However, the relatively short mean
free path and a coherence length which at T=100 mK extends over the
full nanowire, together with the radial confinement, cause a varying
density of states in the leads which leads to these
resonances. This is also consistent with the observation of negative
differential conductance lines. 

%figure2
\begin{figure}[t!]
\includegraphics[width=8.0cm]{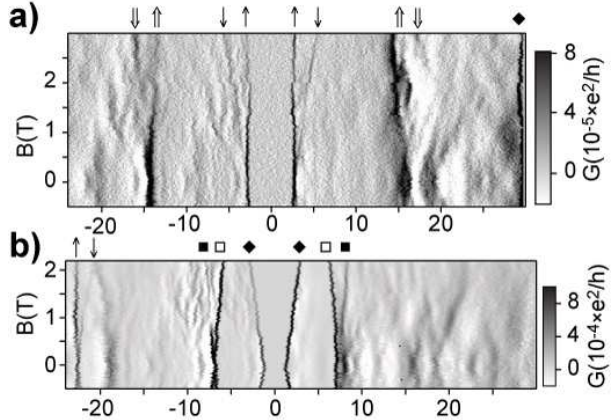}
\caption{\label{fig2} Evolution of the differential conductance peaks
  as a function of $B$ along the two cuts (a) and (b) in
  Fig. \ref{diamond}(b).
} 
\end{figure}
To separate these lead effects from the quantum dot states we
apply a magnetic field $B$ perpendicular to the substrate plane and
measure the differential conductance along two cuts in
Fig. \ref{diamond}(b) detecting the ground-and excited state lines of
the quantum dot with one $(a)$ and two $(b)$ electrons. 
For the one-electron case the peaks in Fig.\;\ref{fig2}(a) on either
side of zero bias split with increasing $B$, reflecting transport
through the Zeeman-split first ($\uparrow,\downarrow$) and second
($\Uparrow,\Downarrow$) orbital level. 
In low-dimensional systems the electronic g-factor $g^*$ depends on system size and dimensionality. It is often found to be strongly reduced from the bulk modulus, which for InAs is $|g^*|\approx 14.7$. \cite{Kiselev98,05bjork} From the Zeeman splitting of orbitals 1 and 2, we extract the effective g-factors $|g^*_{1}|=8\pm0.4$ and $|g^*_{2}|=8.9\pm1$. As expected these values are smaller than the bulk modulus, but consistent with previous experiments performed on similar nanowires.\cite{05bjork}

In Fig.\;\ref{fig2}(b) we identify the two-electron ground state peak ({\footnotesize$\blacklozenge$})
 as the transition from a singlet $S$ to spin-up which involves
 tunneling of a $\downarrow$-electron and therefore moves to higher
 energy with increasing $B$. For the three triplet excited states
 $T^+$,$T^0$,$T^-$ we find two lines
 ({\tiny$\square$}/{\tiny$\blacksquare$}) involving transitions $T^+
 \rightarrow \uparrow$ and $T^0 \rightarrow \downarrow$
 ({\tiny$\square$}) or transitions $T^0 \rightarrow \uparrow$ and $T^-
 \rightarrow \downarrow$ ({\tiny$\blacksquare$}). Here the first two
 transitions ({\tiny$\square$}) require tunneling of a
 $\uparrow$-electron and the transition line moves to lower energy
 with increasing $B$ while in the other two cases
 ({\tiny$\blacksquare$}) a $\downarrow$-electron tunnels and the line
 shows a slope similar to that of the ground state line. In the
 following we refer to the two excited state lines as $\mu_{T^0}$ and
 $\mu_{T^+}$ corresponding to the two dot states carrying most of the
 current.\cite{comm1} The splitting $\mu_{T^0}-\mu_{T^+}$ allows us to extract $|g^*_T|=8\pm 0.5$ similar to the one-electron values. In quantum dots defined in GaAs heterostructures this splitting is typically only observed when a large $B$ is applied in-plane with the 2DEG since the $g$-factor in these systems is much smaller and orbital effects dominate for perpendicular fields. 

A quantitative analysis of our data takes into account asymmetries in
the capacitive coupling from source and drain to the quantum dot. The
conductance peak slopes in Fig. \ref{diamond}(b) change with bias and
plunger gate voltage and differ between states. This indicates that the relation between $V_{G}$, $V_{ds}$ and the energy levels of the quantum dot is non-linear.  We therefore allow the gate leverarm $\alpha_G=C_G/C_{\Sigma}$ to depend on $V_{G}$ but not on $V_{ds}$ and the source/drain coupling asymmetry $\delta\alpha=\alpha_S-\alpha_D$ to vary with $V_{ds}$, where $\alpha_{S(D)}=C_{S(D)}/C_{\Sigma}$ is the source (drain) leverarm.\cite{Ihn04} 
From the two diamond border slopes at each degeneracy point
($V_{ds}\approx0$) we extract $\alpha_G$'s which can be approximated
as $\alpha_G(V_{G})=0.32-(V_{G}+0.905 \mbox{V})/2.27$V. For symmetric
coupling of the dot to source and drain it follows
$\delta\alpha=0$. Using $\alpha_G(V_{G})$, we re-scale
Fig.\;\ref{diamond}(b) eliminating the dependence of borderline slopes
on $V_{G}$. The remaining variations in the slopes are then attributed
to a $V_{ds}$-dependent $\delta\alpha$ which can be determined from
these same slopes. For non-zero $\delta\alpha$, the energy shift of the quantum dot levels with $V_{ds}$ relative to source/drain is given by $\left(1\mp\delta\alpha(V_{ds})\right)V_{ds}/2$ respectively.  

%figure3
\begin{figure}[t!]
\includegraphics[width=8.8cm]{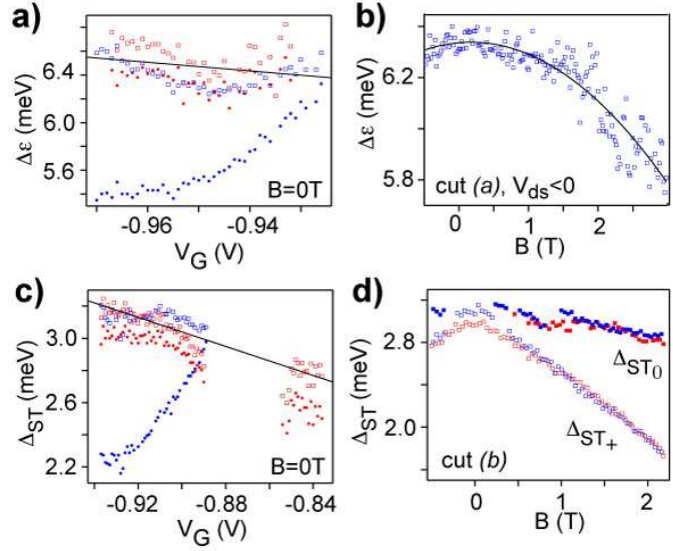}
\caption{\label{Fig3} Orbital splitting $\Delta\varepsilon$ and
  singlet-triplet splitting $\Delta_{ST}$ as a function of $V_G$ and
  $B$. In (a) and (b) $\Delta\varepsilon$ was derived from the
  corrected difference between $\uparrow$ and $\Uparrow$ [see
  Fig.\ref{diamond}(b) and Fig.\ref{fig2}(a)]. 
In (c) $\Delta_{ST}$ was calculated from the data in Fig.\ref{diamond} \emph{cut (a)} and in (d) both splittings $\Delta_{ST^+}$ ($\square$) and $\Delta_{ST^0}$($\blacksquare$) were extracted from Fig.\ref{fig2}(b). For comparison the small dots in panels (a) and (c) indicate the bare bias splittings for negative (blue) and positive (red) $V_{ds}$.}
\end{figure}

The necessity of this correction scheme becomes clear when we extract
the single-particle orbital splitting $\Delta
\varepsilon=\varepsilon_2-\varepsilon_1$ from Fig.\;\ref{diamond}(b)
and plot the bare bias splitting for negative (blue dots)
and positive (red dots) $V_{ds}$ as a function of $V_{G}$ in
Fig.\;\ref{Fig3}(a). The splittings clearly deviate at most $V_{G}$'s,
but after the correction is applied
the splittings for negative (blue squares) and positive (red squares)
$V_{ds}$ are nearly identical and we
find $\Delta \varepsilon\approx6.5\;$meV, changing by less than 4 \%
in the measured interval. The magnetic field dependence of $\Delta\varepsilon$ is extracted from Fig.\;\ref{fig2}(a) and we find a quadratically decreasing behavior [see Fig. \ref{Fig3}(b)]. This is similar to previous results in quantum dots with a 2D harmonic confinement.\cite{Tarucha96, Ciorga00} For comparison, we apply a similar harmonic confinement model\cite{85schuh,suppl} and extract $\hbar\omega_x=6.3\;$meV and $\hbar\omega_y=40\;$meV from the fit indicated by the black line. This is in agreement with the device geometry from which we expect that the dot is elongated along the nanowire. We further estimate an effective size $2\lambda_x=2\sqrt{\hbar/m^*\omega_x}=46\;$nm and $2\lambda_y=18\;$nm.

For the two-electron case we determine the singlet-triplet splitting at $B=0$ with $\Delta_{ST}=\mu_{T}-\mu_{S}$ [see Fig.\;\ref{Fig3}(c)]. Again the two bias directions give the same result after the correction has been applied and $\Delta_{ST}$ decreases from 3.2\;meV to 2.8\;meV over the measured gate voltage range. The opposite dependence of $\Delta_{ST}$ on plunger gate voltage was previously observed in a lateral gate-defined quantum dot in a GaAs based two dimensional electron gas and attributed to shape deformation of the parabolic potential.\cite{Kyriakidis02,Zumbuhl04} 

For the magnetic field dependence of $\Delta_{ST}$ [see
Fig.\;\ref{Fig3}(d)] we consider two splittings
$\Delta_{ST^+}=\mu_{2,T^+}-\mu_{2,S}$ and
$\Delta_{ST^0}=\mu_{2,T^0}-\mu_{2,S}$. $\Delta_{ST^0}$ only shows a
small shift with $B$, which we attribute to orbital and interaction
effects. Due to the large Zeeman splitting between the triplet states,
$\Delta_{ST^+}$, shows a strong linear decrease
with increasing $B$. In addition, we have extracted the Coulomb interaction energy $U = 13.5$\; meV between two electrons on the first orbital using the corresponding conductance peak splitting in Fig.\;\ref{diamond}(b). We found $U$ to be independent of $V_{G}$ and $B$ within the investigated ranges.

%figure4
\begin{figure}[t!]
\includegraphics[width=7.8cm]{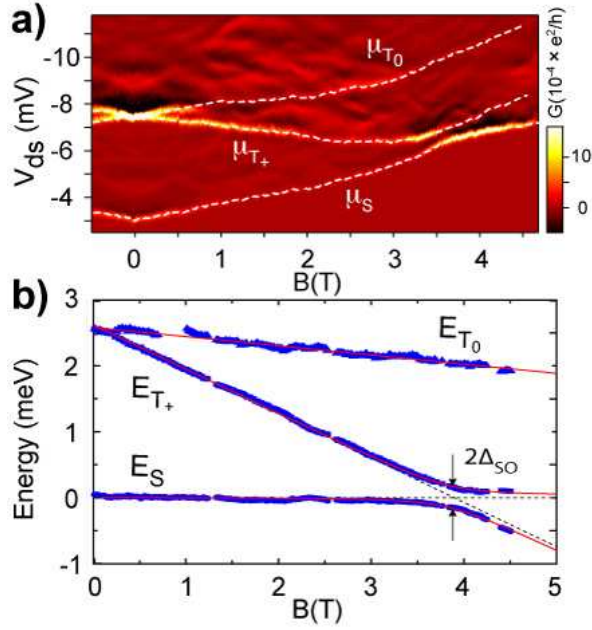}
\caption{\label{Fig4} (a) Singlet-triplet transition from cut (b) in
  Fig. \ref{diamond} after gate voltages have been optimized. The
  transition between $\mu_{T^+}$ and $\mu_{S}$ exhibits a clear
  anti-crossing. (b) Energy levels extracted from the corrected peak
  splittings in (a) together with a fit to a simple model indicated by
  the red lines. 
} 
\end{figure}
Figure\;\ref{Fig4}(a) again gives the magnetic field dependence of
$\mu_T$ and $\mu_S$ similar to that in Fig. \ref{fig2}(b) but over an
extended magnetic field range. 
At $B\approx3.8\;$T the quantum dot undergoes a transition where the two-electron groundstate changes from $S$ to $T^+$.  In contrast to the usually observed crossing\cite{Kouwenhoven01,Zumbuhl04,Ellenberger06}, the $S-T$ transition in our data clearly shows an avoided crossing due to the spin-orbit interaction. We then extract the separation of these three chemical potentials and by assuming that the singlet energy $E_S$ is independent of $B$ we can reconstruct the spectrum of the two-electron artificial atom as shown in Fig.\;\ref{Fig4}(b). The magnitude of the anti-crossing is $\Delta_{SO}=0.23\;$meV and serves as a direct measure of the spin-orbit interaction strength in such quantum dots.

The InAs nanowires are known to have wurtzite crystal
structure\cite{05bjork}, despite the fact that bulk wurtzite InAs does
not exist. The wurtzite crystal has a reduced symmetry compared to
zinc-blende structures and because of $s$-$p_z$ mixing, additional spin-orbit terms linear in
momentum occur for the conduction band electrons in bulk materials. In
nanowires, like in most nanosystems, an important role is played by
the boundary and additional spin-orbit terms arise in asymmetric
structures, such as ours. Generally the spin couples to the longitudinal motion of the electron with a spin-orbit interaction that in first order of ${\bm k}\cdot{\bm c}$ assumes the following general form
\begin{equation}
H_{SO}=({\bm k}\cdot{\bm c})(\mbox{\boldmath $\gamma$}\cdot\mbox{\boldmath $\sigma$}),
\label{eqHS0gamma}
\end{equation}
where ${\bm c}$ is a unit vector along the wurtzite hexagonal axis ($c$ axis) 
and $\mbox{\boldmath $\gamma$}=(\gamma_x,\gamma_y,\gamma_z)$ is a vector of coupling constants. In a Hund-Mulliken approximation for the elongated nanowire quantum dot this general form of the spin-orbit interaction leads to efficient mixing of the $S$ and $T^+$ states with the magnitude of the anti-crossing given by
\begin{equation}
\Delta_{SO}(B)=\frac{E_Z }{\sqrt{2}}\frac{r_{12}\tilde m_\|^*}{\hbar^2}\sqrt{\gamma_x^2+\gamma_y^2}=\frac{E_Z }{\sqrt{2}}\frac{r_{12}}{\lambda_\mathrm{SO}},
\label{eqDelta}
\end{equation}
where $E_Z=g^*\mu_BB$ is the Zeeman energy, $r_{12}=\left\langle\psi_T|(x_1-x_2)|\psi_S\right\rangle$ 
is an effective distance between the two electrons, and $\tilde m^*_\|$ is the electron effective mass in the nanowire
in a magnetic field. 
The red lines in Fig.\;\ref{Fig4}(b) are a fit to the standard expression for level repulsion 
\begin{equation}
E_{1,2}=\frac{E_S+E_{T_+}}{2}\pm\sqrt{\frac{\left(E_S-E_{T_+}\right)^2}{4}+\Delta_{SO}^2},
\label{supE12}
\end{equation}
where the bare singlet energy was set to be constant $E_S = 0\;$ and $E_{T_+}=E_{T_0}-g^*\mu_BB$ with $E_{T_0}=-\gamma_B B$. We assume that $r_{12}$ does not depend on $B$. This means that $\Delta_{SO}$ increases linearly with magnetic field and yields an asymmetric splitting when comparing the left with the right hand side of the anticrossing. The linear dependence of $E_{T_0}$ on magnetic field is an experimental finding consistent with all acquired datasets and characterised by the phenomenological parameter $\gamma_B=0.14\;$meV/Tesla. At $B\approx3.8\;$T where the bare levels cross (dashed lines) we find $E_Z\approx 1.8\;$meV and estimate $r_{12}\approx\lambda_x=23\;$nm which allows us to calculate the spin-orbit length $\lambda_{SO}\approx 127\;$nm from Eq.~(\ref {eqDelta}). This value is comparable to $\tilde{\lambda}_{SO}\approx200\;$nm obtained for unconfined electrons in similar InAs nanowires\cite{05hansen}. 

\begin{acknowledgments}
This work was supported by the Swedish Foundation for Strategic Research (SSF), the Swedish Research Council (VR) and the Swiss NSF, NCCR Nanoscience Basel, and ICORP JST (VG \& DL).
\end{acknowledgments}

\bibliographystyle{apsrev}

\newpage
\setcounter{page}{1}
\subsection*{{\Large Supplementary Notes}}
\newcounter{Lcount}
\vskip 2cm
\begin{list}{\bf\Alph{Lcount}.}
    {\usecounter{Lcount}}
    \item {\bf Detecting the Last Electron}
    \item {\bf $g$-factor and singlet-triplet anti-crossing for different dot configurations}
    \item {\bf B-field dependence of the bare energy levels of an elongated\\ quantum dot with harmonic confinement}
    \item {\bf Origin of the SO-interaction in wurtzite InAs and\\ avoided crossing between $S$ and $T^+$}
\end{list}

\date{\today}

\newpage
\subsection*{A. Detecting the Last Electron}

\begin{figure}[b!]
\begin{center}
\includegraphics[width=15cm]{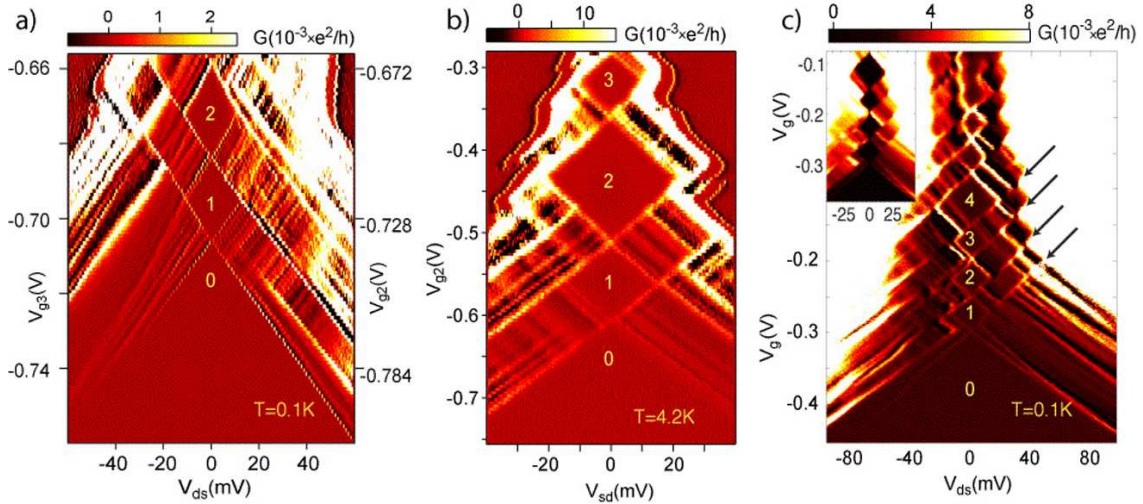}
\end{center}
\caption{\label{Fig1s} (a) Differential conductance for a dot with similar settings as in the main text over a larger bias range ($V_{g1}=-2.274$V, $V_{g4}=-1.630$V and $V_{g5}=$2.2V).  (b) Differential conductance for a dot confined
  by gates 1 and 3 ($V_{g1}=-1.5$V, $V_{g3}=-0.7$V, $V_{g4}$ and
 $V_{g5}$ are both at -0.2V) measured with gate 2 as
  plunger gate at 4.2 K. (c) Main figure: Differential conductance of a second sample with a similar quantum dot. Arrows indicate alignment of the chemical potential of the dot with that of the drain. Inset: absolute value of the current trough the same device. }
\end{figure}

To support our statement that it is possible to empty the dot completely of
electrons, we have measured Coulomb blockade diamonds on two different samples and in different cool-downs. Fig.\ref{Fig1s}(a) shows the differential
conductance in the few-electron regime from a second cool-down in the dilution refrigerator using a configuration similar to the one discussed in the main text. Gates 2 and 3 are again used
simultaneously as plunger gates, but we extended the sweeps to larger bias voltages in order to show the opening of the $N$ = 0 region without any features that would indicate further electrons leaving the dot. In all the measurements with two plunger gates the relative lever-arm between them is $C_{g2}/C_{g3}=0.85\pm 0.05$. This shows that the quantum dot in the nanowire is similarly coupled and tuned by both plunger gates and therefore likely to be centred in the middle. Figure \ref{Fig1s}(b) was measured at 4.2 K for the same sample. 

Fig.\ref{Fig1s}(c) shows Coulomb diamonds on a second sample with a similar nanowire in a 3-gate configuration. Here, the last electron leaves the quantum dot at $V_\mathrm{g}=-0.3$\;V. This is supported by the fact that no further borderlines (see black arrows) appear at positive bias after the chemical potential of the drain contact has dropped below the chemical potential of the last electron on the dot (lowest arrow). In other samples [see main text or Fig.\ref{Fig1s}(a) and (b) ] similar but less distinct lines (both differential and negative differential conductance lines) \emph{do} occur even below the N=1 diamond. This is something that we observe in many of our nanowire based quantum dots. Specifically, also in the InP/InAs heterostructure based QD's where the electron number can be determined without any doubt\cite{05bjork}. The origin of these peaks lies in the fact that our leads to the QD's are nearly one-dimensional (in contrast to 2DEG based dots). A relatively short mean free path and a coherence length at T = 100 mK which extends over the full nanowire, lead to the observation of resonances that have their origin in the leads and not the quantum dot. Such an explanation is also consistent with the observation of negative differential conductance lines. We use the magnetic field dependence of the differential conductance peaks to determine which peaks belong to the dot and which ones originate in the leads. The dot states typically exhibit a "slow" variation as a function of magnetic field while the leads states vary quickly in energy with B and tend to be suppressed for large fields [ see e.g. Fig. 2 (b), (c) and Fig. 3 (b), (c) ]. In addition, by tuning the gates which lie below the contacts (left of gate 1 and right of gate 4) we were able to tune the lead states while leaving the dot related peaks intact, therefore excluding the possibility that these additional resonance lines originate from low lying excited states in the dot. 

We have fabricated and performed measurements at low temperatures on six devices with a gate period of 60\;nm. Two of these had five or more working gates without any leakage or connection problems and in both we were able to access the last electron in a gate defined quantum dot.

\newpage
\subsection*{B. $g$-factor and singlet-triplet anti-crossing for different dot configurations}
\begin{figure}[b!]
\begin{center}
\includegraphics[width=12cm]{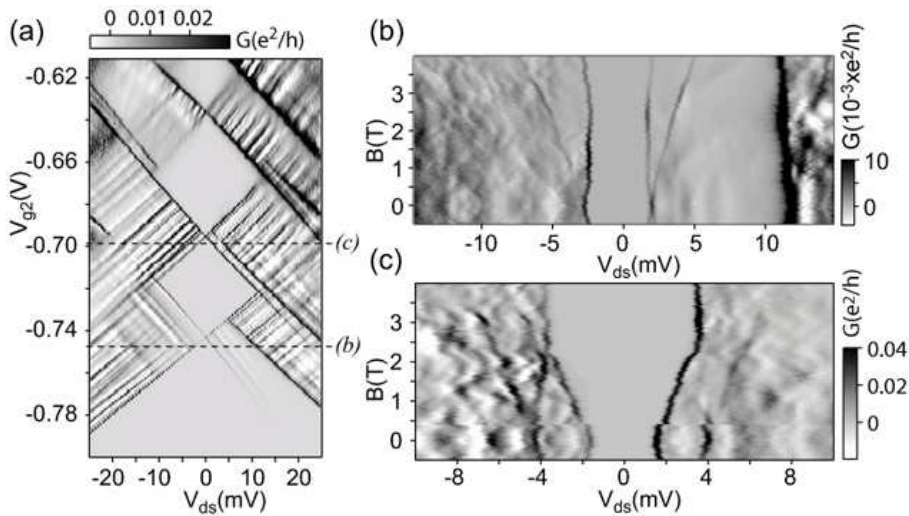}
\caption{\label{Fig2s} (a) Differential conductance measured for a quantum dot
  confined by gates 1 ($V_{g1}=-2.56$V) and 3 ($V_{g3}=-1.72$V). Gates
  4 and 5 are both at 2.5 V and gate 2 is used as the plunger
  gate. (b) and (c) Evolution in magnetic field of the conductance peaks along
  cuts \emph{(b)} and \emph{(c)} in panel (a).  
  }
\end{center}
\end{figure}
\begin{figure}[b!]
\begin{center}
\includegraphics[width=12cm]{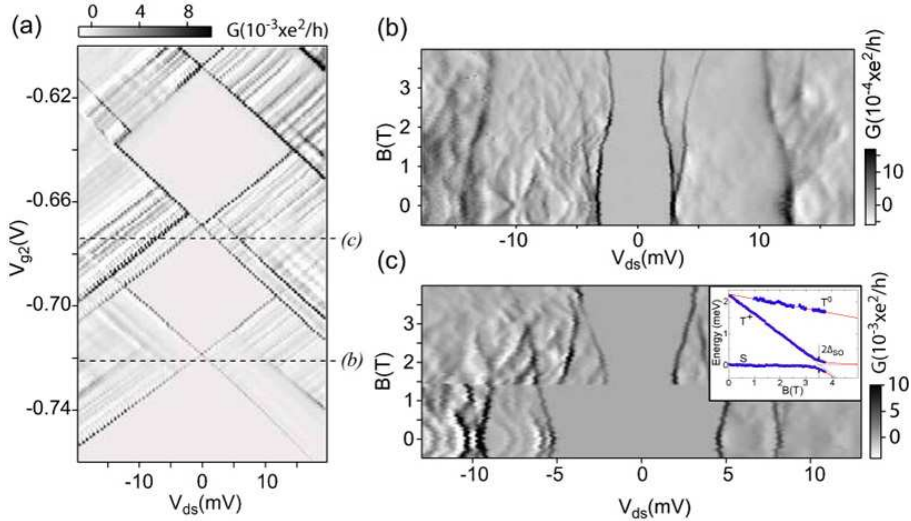}
\caption{\label{Fig3s} (a) Differential conductance measured for a
  quantum dot confined by gates 1 ($V_{g1}=-2.56$V) and 3/4
  ($V_{g3}=-1.17$V and $V_{g4}=-0.5$V). $V_{g5}=2.5$V and gate 2 is
  used as the sole plunger gate. (b) and (c) Evolution in magnetic
  field of the conductance peaks along cuts \emph{(b)} and \emph{(c)}
  in panel (a). The inset in (c) shows the singlet and triplet energies and the anticrossing at $B\approx3.5$\;T.} 
\end{center}
\end{figure}

The current device layout permits the definition of very flexible quantum
dots, where couplings and the quantum dot position and geometry can be varied. In the main text, we consider a fairly extended dot, tuned by double plunger gates (4-gate dot). Due
to the tunability of our device design, we can also define smaller
dots only with a single plunger gate (3-gate dot). As the
quantum dot is tuned between these two configurations, with either one
or two plunger gates, the main features of the differential
conductance as a function of bias and plunger gate voltage are preserved. Most notably the avoided crossing between the singlet and triplet two-electron
states is found for both gate settings. Figs.\ref{Fig2s} and \ref{Fig3s} show similar measurements as Figs. 1 and 2 in the main text, but for two additional gate configurations.

Fig. \ref{Fig2s} shows Coulomb diamonds and magnetic field
dependence of the conductance peaks for a three gate dot with gate
2 acting as the plunger gate. The Zeeman splitting and first orbital excited state as well as the triplet states are clearly identifiable. From the Zeeman splitting of the first and
second orbital states in Fig.\ref{Fig2s}(b), we extract effective
g-factors $|g_1^*|=8.6\pm0.3$ and $|g_2^*|=7.6\pm0.5$, which are again in the same range as for the dot in the main text. The discrepancy in the magnitude of the Zeeman splitting (clearly visible in
Fig.\ref{Fig2s}(b)) is reduced to a negligible difference when the correction scheme described in the main text is applied. In Fig. \ref{Fig2s}(c) we can again identify the transitions from the
single-electron to a singlet or triplet state. At B$=2.5$T the peaks
anti-cross.
Figure \ref{Fig3s} shows the corresponding data for a quantum dot
defined at intermediate gate settings. Compared to
Fig. \ref{Fig2s}, $V_{g3}$ is tuned more positive and $V_{g4}$ more negative,
resulting in a configuration where the right-hand barrier is
shared between gates 3 and 4. From the splitting of the states in Fig. \ref{Fig3s}(b) we can
again extract effective $g$-factors which are found to be
$|g_1^*|=7.7\pm0.5$ and $|g_2^*|=7.6\pm0.5$.

The cut for the two- electron case in Fig.\ref{Fig3s}(c), shows a charge rearrangement at $B=1.5\;$T. Extracting the singlet and triplet energies (see
inset) shows, however, that there has been no change in the relative
positions of the conductance peaks. Again the anti-crossing is
clearly present in this case.

\newpage
\subsection*{C. B-field dependence of the bare energy levels of an elongated quantum dot with harmonic confinement.}
The B-field dependence of the bare energy levels (without spin-orbit interaction) of a quantum dot with harmonic 
confinement in all three directions can be calculated by
diagonalizing a quadratic form~\cite{Schuh}.
For the weakest confinement direction along the axis of the wire ($x$-axis) and a transverse magnetic field
$\mbox{\boldmath $B$}\| \mbox{\boldmath $z$}$, the lowest excitation energy of a single-electron quantum dot is given by
\begin{equation}
\Delta\varepsilon=\frac{\hbar}{2}\left[\sqrt{(\omega_y+\omega_x)^2+\omega_c^2}
-\sqrt{(\omega_y-\omega_x)^2+\omega_c^2}\right]
\label{supdeltaepsilon}
\end{equation}
where $\hbar\omega_x$ and $\hbar\omega_y$ are, respectively, the confinement energies for motion along the wire 
and in the direction transverse to both wire and magnetic field;
$\omega_c=eB/c\sqrt{m^*_\|m^*_\perp}$ is the cyclotron frequency.
For small $\omega_c,\omega_x\ll\omega_y$, one obtains a quadratic B-field dependence of the excitation energy
\begin{equation}
\Delta\varepsilon=\hbar\omega_x\left(1-\frac{\omega_c^2}{2\omega_y^2}\right).
\end{equation}
The values $\hbar\omega_x=6.3\;{\rm meV}$ and $\hbar\omega_y=40\;{\rm meV}$ given in the main text were determined using a fit to the measured data points $\Delta\varepsilon(B)$ in Fig.\;3(b) of the main text.

\newpage
\subsection*{D. Origin of the SO-interaction in wurtzite InAs and avoided crossing between $S$ and $T^+$}

The spin-orbit interaction in Eq.~(1) of the main text 
is peculiar to nanowires in which the transverse quantization
is much stronger than the longitudinal one.
In the absence of magnetic fields, the lowest electron mode in the nanowire is doubly degenerate, due to the
time-reversal symmetry, and thus the dispersion relation obeys
\begin{equation}
\epsilon_{\uparrow}(k_x)=\epsilon_{\downarrow}(-k_x),
\end{equation}
where $\epsilon_{s}(k_x)$ is the energy of the electron with longitudinal momentum $k_x$ 
and spin $s$ quantized along the direction of the internal field.
The spin-splitting of the mode is absent at $k_x=0$ and can be expanded in powers of $k_x$
\begin{equation}
\frac{\epsilon_\uparrow(k_x)-\epsilon_\downarrow(k_x)}{2}= \gamma k_x+ {\cal O}(k_x^3).
\label{SpinSplitting}
\end{equation}
In Eq.~(1) of the main text, we retain, for simplicity, only the linear in $k_x$ terms.
Inclusion of an arbitrary odd function of $k_x$
does not change our results substantially, while it can be viewed as redefining
the $\gamma$'s in Eq.~(2) of the main text. 

The linear in $k$ terms in Eq.~(1) of the main text are
different from the Rashba spin-orbit interaction of the bulk wurtzite~\cite{60rashba,Casella}.
The latter has the form $H_{SO}^{(1)}=C[\mbox{\boldmath $k$}\times\mbox{\boldmath $c$}]\cdot\mbox{\boldmath $\sigma$}$,
where ${\bm c}$ is a unit vector along the c-axis of the wurtzite (uniaxial) crystal.
In the case of wurtzite InAs nanowires, ${\bm c}$ is also the unit vector along the wire.
Unlike in zinc-blende alloys, the Rashba terms are present in wurtzite in the bulk
and values of $C\simeq(0.1-10)\times 10^{-10}\,{\rm eV}\,{\rm cm}$ have been reported for A$_{\rm II}$B$_{\rm VI}$ 
compounds~\cite{Hopfield,Ohta,LewYanVoon}. 
For wurtzite InAs, the value of $C$ is not known.

Despite the fact that the Rashba terms, if present, might couple the spin efficiently to the transverse electron motion,
these terms do not mix singlet and triplet states in nanowires with soft confinement along ${\bm c}$.
Notably, after averaging over the transverse electron motion the Rashba terms vanish.
We have further verified that $H_{SO}^{(1)}$ gives no contribution to $\Delta_{SO}$ up to the third order of 
perturbation theory in $H_{SO}^{(1)}$, provided the quantum dot electron 
has well separated scales of a tight transverse and extended
longitudinal motion. In the opposite limiting case (not realized in our experiment), 
when the gate-defined confinement along the wire is stronger than
the transverse quantization, we expect $\Delta_{SO}$ to be a measure of $H_{SO}^{(1)}$.

Similar to zinc-blende compounds, 
third and higher (odd) powers of ${\bm k}$ also contribute to the spin-splitting in the conduction band of wurtzite. 
We denote these terms by $H_{SO}^{(2)}=\mbox{\boldmath $h$}(\mbox{\boldmath $k$})\cdot\mbox{\boldmath $\sigma$}$, 
where $\mbox{\boldmath $h$}(\mbox{\boldmath $k$})$ obeys
$\mbox{\boldmath $h$}(k\mbox{\boldmath $c$})=0$, $\forall k$, 
due to the uniaxial symmetry of the wurtzite~\cite{Casella}.
At the third order of ${\bm k}$, $\mbox{\boldmath $h$}(\mbox{\boldmath $k$})$ 
reads~\cite{Zorkani}
$\mbox{\boldmath $h$}(\mbox{\boldmath $k$})=\left(\lambda_l(\mbox{\boldmath $k$}\cdot\mbox{\boldmath $c$})^2
+\lambda_t(\mbox{\boldmath $k$}\times\mbox{\boldmath $c$})^2\right)
\left[\mbox{\boldmath $k$}\times\mbox{\boldmath $c$}\right]$,
where $\lambda_l$ and $\lambda_t$ are coupling constants.
At higher orders in ${\bm k}$, we expect $\mbox{\boldmath $h$}(\mbox{\boldmath $k$})$
to give a finite contribution to Eq.~(1) of the main text,
after averaging over the transverse electron motion.

Apart from the crystalline inversion asymmetry, an important source of spin-orbit
interaction in nanowires is the structure inversion asymmetry at the radial edge of the sample.
This mechanism is analogous to the Bychkov-Rashba spin-orbit interaction in heterostructures~\cite{84bychkov}
and in quantum wires it gives a finite contribution to Eq.~(1), provided the electron
is displaced from the center of the wire to one of its edges. 
Due to the particular way of gating our quantum dots (with gates beneath the wire) we expect
a strong contribution from this mechanism in our specific case.
Additional possible sources of spin-orbit interaction in nanowires
include built-in radial strain, charged surface states, deviations of the InAs lattice
from wurtzite (e.g. occurrences of the zinc-blende stacking sequence along ${\bm c}$), etc.

Next we show that the spin-orbit interaction in Eq.~(1) of the main text
mixes the singlet ($S$) and triplet ($T^+$) states and derive Eq.~(2) of the main text.
Implying the Hund-Mulliken approximation, we write the singlet and triplet orbital wave functions 
as follows 
\begin{equation}
\psi_S({\bm r}_1,{\bm r}_2)=\frac{1}{\sqrt{1+\phi^2}}\left(\psi_+({\bm r}_1)\psi_+({\bm r}_2)-
\phi\psi_-({\bm r}_1)\psi_-({\bm r}_2)\right),
\end{equation}
\begin{equation}
\psi_T({\bm r}_1,{\bm r}_2)=\frac{1}{\sqrt{2}}\left(\psi_+({\bm r}_1)\psi_-({\bm r}_2)-
\psi_-({\bm r}_1)\psi_+({\bm r}_2)\right),
\end{equation}
where $\phi$ is the so-called interaction parameter~\cite{BLD,GLdd,Zumbuel} and 
$\psi_\pm({\bm r})=\exp(iM)f_\pm(x)g(y,z)$, with $f_\pm(x)=\pm f_\pm(-x)$ being the trial wave functions
that minimize the energy of the two interacting electrons in the quantum dot.
Here, we choose $x$ parallel to the nanowire and $z$ along the magnetic field.
The orbital effect of the magnetic field can be accounted for by the operator $M$,
which assumes $M=\frac{\hbar eB}{m^*_\|m^*_\perp\omega_y^2}\frac{\partial^2}{\partial x\partial y}+{\cal O}(B^2)$
in the Landau gauge ${\bm A}({\bm r})=(-y B,0,0)$.
$M$ mixes the wave function components $g(y,z)$ and $f_\pm(x)$ to an
amount controlled by the transverse oscillator frequency $\omega_y$.
However, in our further derivation the orbital magnetic field is not 
essential and we retain it only for the sake of completeness.

Considering further a generic Zeeman field, 
\begin{equation}
H_Z=\frac{1}{2}E_Z{\bm n}\cdot\mbox{\boldmath $\sigma$},
\end{equation}
where $E_Z$ is the Zeeman splitting and ${\bm n}$ is the spin quantization axis,
we transform away the spin-orbit interaction obtaining 
\begin{equation}
H_{SO}+H_Z\to H_Z+H_Z^{SO},
\end{equation}
where $H_Z^{SO}=E_Z(\tilde m_\|^*/\hbar^2)({\bm r}\cdot{\bm c})\left[{\bm n}\times{\bm \gamma}\right]\cdot\mbox{\boldmath $\sigma$}$ 
gives the combined effect the spin-orbit and Zeeman interactions.
The longitudinal effective mass is slightly larger due to the orbital magnetic field effect,
\begin{equation}
\frac{1}{\tilde m_\|^*}=\frac{1}{m^*_\|}\left[1-\frac{\omega_c^2}{2\omega_y^2+\omega_c^2}\right].
\end{equation}
However, using $\hbar\omega_y=40$\;meV and for $B<5$\;Tesla this correction is negligible and we assume 
$\tilde m_\|^*=m^*_\|=m^*_\perp=m^*_{bulk}=0.023\;m_e$ for all numerical estimates.

Taking matrix elements of $H_Z^{SO}$ for two electrons and between the singlet and triplet states,
we obtain the splitting $2\Delta_{SO}$, with
\begin{equation}
\Delta_{SO}=
\left|\langle\Psi_S| H_Z^{SO}|\Psi_{T^+}\rangle\right|
=E_Zr_{12}\frac{\tilde m_\|^*}{\sqrt{2}\hbar^2}\sqrt{\gamma^2-
(\mbox{\boldmath $\gamma$}\cdot\mbox{\boldmath $n$})^2}.
\label{supDeltaSO}
\end{equation}
Next we evaluate the distance $r_{12}=\left\langle\psi_T|(x_1-x_2)|\psi_S\right\rangle$
in two limiting cases: (i) strong Coulomb interaction and (ii) weak Coulomb interaction.
For our quantum dots, case (ii) turns out to be relevant.

For strong Coulomb repulsion, we construct $f_\pm(x)$ from
two Gaussian wave packets centered around $x=\pm d/2$ and with
overlap ${\cal S}$. Then we obtain
\begin{equation}
r_{12}=\frac{1+\phi}{\sqrt{2(1+\phi^2)}}\frac{d}{\sqrt{1-{\cal S}^2}}.
\end{equation}
A starting approximation for $d$ is found by minimizing the potential energy
of the two electrons, $W=e^2/\kappa d+m^*_\|\omega_x^2d^2/4$.
The result reads $d=\lambda_x\left(2\lambda_x/a_B\right)^{1/3}\approx25$\;nm,
where $a^*_B=\hbar^2\kappa/m^*_\|e^2$ is the Bohr radius, with $\kappa\approx15$ being the dielectric constant of bulk InAs,
and $\lambda_x=\sqrt{\hbar/m^*_\|\omega_x}=23$\;nm the effective longitudinal dot size.
Similarly, the width of the Gaussian wave packet $\lambda_0$ can be estimated by minimizing the sum of the 
confining and exchange energies, 
${\cal E}=\hbar^2/2\tilde m_\|^*\lambda_0^2+{\cal S}^2e^2/\kappa d$, where ${\cal S}=\exp(-d^2/4\lambda_0^2)$.
The result for $\lambda_0$ is then given by $\lambda_0=d\left[2\ln(d/\tilde a_B^*)\right]^{-1/2}$, 
where $\tilde a_B^*=\hbar^2\kappa/m_\|^*e^2$ is the effective Bohr radius.
Similarly, we find ${\cal S}=\sqrt{\tilde a_B^*/d}$.
In the strong Coulomb interaction regime, 
defined by $\lambda_x/a^*_B\gg 1$ and $d/\tilde a_B^*\gg 1$, 
one can further evaluate $\phi$ with the help of the expression 
$\phi=\sqrt{1+(4t_H/U_H)^2}-4t_H/U_H$ and by letting
$t_H\simeq{\cal S}\hbar^2/\tilde m_\|^*\lambda_0^2$ and $U_H\simeq e^2/\kappa\lambda_0$.

In the absence of Coulomb interaction, we obtain
\begin{equation}
r_{12}=\sqrt{\frac{\hbar}{2m^*_\|\omega_x}}\left[1-\frac{\omega_c^2}{2\omega_y^2+\omega_c^2}\right]^{1/4}\approx\lambda_x.
\end{equation}
We note that this expression gives a lower bound for $r_{12}$ in the presence of Coulomb interaction.


\begin{thebibliography}{31}
\bibitem[{com()}]{comm1}
\bibinfo{howpublished}{The two processes $T^-\rightarrow\downarrow$ and
  $T^0\rightarrow\downarrow$ also contribute to the current but are less
  probable and found to be suppressed in our experiment. This is likely due to
  fast spin relaxation combined with weak coupling to the leads. The process
  $T^-\rightarrow\uparrow$ is forbidden by selection rules.}
\end{thebibliography}
\end{document}